\newcommand{\setup}[1]{%
    \ifcase#1\relax
    \documentclass{article}
    \usepackage[margin=1in]{geometry}
    \renewcommand{\baselinestretch}{2}
    \bibliographystyle{Chicago}
    \or 
    \documentclass{scrartcl}
    \bibliographystyle{unsrtnat}
    \fi
}
\setup{1}

\usepackage[utf8]{inputenc}
\usepackage{amsmath}
\usepackage{amssymb}
\usepackage{authblk}
\usepackage[colorlinks=true,allcolors=blue]{hyperref}
\usepackage{amsthm}
\usepackage{cleveref}
\usepackage{booktabs}
\usepackage{graphicx}
\usepackage{verbatim}
\usepackage[giveninits=true,sorting=none]{biblatex}
\renewbibmacro{in:}{}
\DeclareFieldFormat{pages}{#1}
\DeclareMathOperator{\var}{var}
\DeclareMathOperator{\cov}{cov}

\newcommand{\E}[1]{\left\langle#1\right\rangle}

\newcommand{\dist}{\sim}
\newcommand{\abs}[1]{\left\vert#1\right\vert}
\newcommand{\real}{\mathbb{R}}
\newcommand{\prob}[1]{p\left(#1\right)}

\newcommand{\Normal}{\mathsf{Normal}}
\newcommand{\BetaBinomial}{\mathsf{BetaBinomial}}
\newcommand{\citet}[1]{\textcite{#1}}

\usepackage{xurl}

\begin{document}
\title{Latent space approaches to aggregate network data}
\author{Till Hoffmann}
\affil{\small Department of Mathematics, Imperial College London\\\href{mailto:t.hoffmann@imperial.ac.uk}{t.hoffmann@imperial.ac.uk}}
\date{}
\maketitle

\begin{abstract}
    Large-scale network data can pose computational challenges, be expensive to acquire, and compromise the privacy of individuals in social networks. We show that the locations and scales of latent space cluster models can be inferred from the number of connections between groups alone. We demonstrate this modelling approach using synthetic data and apply it to friendships between students collected as part of the Add Health study, eliminating the need for node-level connection data. The method thus protects the privacy of individuals and simplifies data sharing. It also offers performance advantages over node-level latent space models because the computational cost scales with the number of clusters rather than the number of nodes.
\end{abstract}

\section{Introduction}

Friendships between people, academic co-authorships, and protein-protein interactions can all be modelled by networks~\cite{Newman2018}---a powerful paradigm for investigating connections between different entities. Latent space approaches model the probability for two entities to be connected as a function of the distance between them in an unobserved space~\cite{Hoff2002}. This latent space offers an intuitive interpretation of the relations between entities, and the modelling approach can account for important aspects of real-world networks, such as transitivity and reciprocity~\cite{Rastelli2016}. The models can be easily fit to network data~\cite{Hoff2002,Raftery2012,Salter-Townshend2013}, but when these data comprise connections between individuals they can pose a significant privacy risk~\cite{Lazer2021}. For example, sociotechnical systems, such as telecommunication infrastructure and online social networks, offer a ``telescope'' to observe human behaviour at unprecedented granularity~\cite{Golder2014}. The wealth of data available about individuals can be used to re-identify them in large, supposedly anonymised datasets~\cite{Backstrom2007} or even to infer sensitive attributes~\cite{Kosinski2013}. To protect the privacy of individuals, access to these data is often restricted, creating a ``digital divide'' between the ``data rich'' with privileged access and the ``data poor'' without~\cite{Boyd2012}. This divide not only hinders scientific progress but also poses challenges for reproducibility because findings cannot be reproduced without access to the data.

What if we could fit latent space models to data without ever having to access sensitive, individual-level information? Aggregating data to obtain a ``coarse grained'' dataset is an established approach for protecting privacy while still capturing information about all individuals~\cite{Cox1980}. Recent advances have shown that it is possible to ``identify network structure without network data'' by modelling the number of connections between an individual and a group, such as people sharing the same name or ethnicity~\cite{McCormick2015,Breza2020}. Yet models for \emph{fully aggregated} network data, i.e. a dataset comprising only connection volumes between groups, are lacking. 

In this paper, we develop methods to fit latent space models to fully aggregated network data in \cref{sec:methods}, and we show that latent properties can be inferred from both synthetic and real-world data given only connection volumes between groups in \cref{sec:results}. This approach not only protects the privacy of individuals but also offers performance benefits because the computational cost no longer scales with the number of nodes in the network but rather with the number of groups. In \cref{sec:discussion}, we discuss limitations due to using only aggregate network data and propose next steps for refining the inference methodology.

\section{Methods \label{sec:methods}}
\subsection{Latent space models \label{sec:model}}

Latent space models endow each node $i$ in a population of $n$ nodes with coordinates $z_i\in \real^q$ in a $q$-dimensional latent space~\cite{Hoff2002}. The nodes' positions in the latent space encode how they relate to one another. We use an adjacency matrix $y$ to represent connections between nodes such that $y_{ij}=1$ if node $j$ is connected to node $i$ and $y_{ij}=0$ otherwise. The probability for nodes $i$ and $j$ to connect is
\[
    \prob{y_{ij}=1\mid z_i,z_j,\theta} = \lambda\left(z_i,z_j,\theta\right),
\]
where $\theta$ parameterises the connectivity kernel $\lambda$, which typically decays with increasing separation between the nodes. Connections are conditionally independent given the latent coordinates $z$. The kernel is assumed to be symmetric with respect to exchange of coordinates, i.e. $\lambda\left(z_i,z_j,\theta\right)=\lambda\left(z_j,z_i,\theta\right)$. We consider unweighted, directed networks, but subsequent results generalise to weighted and undirected networks (see \cref{app:statistics} for details). 

Each node $i$ belongs to one of $r$ groups $g_i$, and members of the same group have a common coordinate distribution in the latent space, i.e.
\[
    z_i\mid\mu,\sigma,g_i \dist\Normal\left(\mu_{g_i},\sigma^2_{g_i}\right),
\]
where $\mu_g\in\real^q$ is the centre of group $g$ and $\sigma^2_g>0$ is its variance. This setup closely follows the model-based clustering approach for latent space models developed by \citet{Handcock2007}. 

In contrast to latent space cluster models~\cite{Handcock2007} and community detection algorithms~\cite{Fortunato2010}, which are designed to assign nodes to latent groups given the individual-level network $y$, we assume that groups are defined a priori and that individual connections remain hidden. For example, groups could be defined by members having certain properties in common, such as sex, age, physical location, or a combination thereof. Only aggregate connections $Y$ between groups are observable. In particular, the number of connections from group $b$ to group $a$ is
\begin{equation}
    Y_{ab}=\sum_{i\neq j} \delta_{ag_i}\delta_{bg_j}y_{ij},
    \label{eq:aggregation-definition}
\end{equation}
where $\delta_{ab}$ is the Kronecker delta.

\begin{figure}
    \centering
    \includegraphics{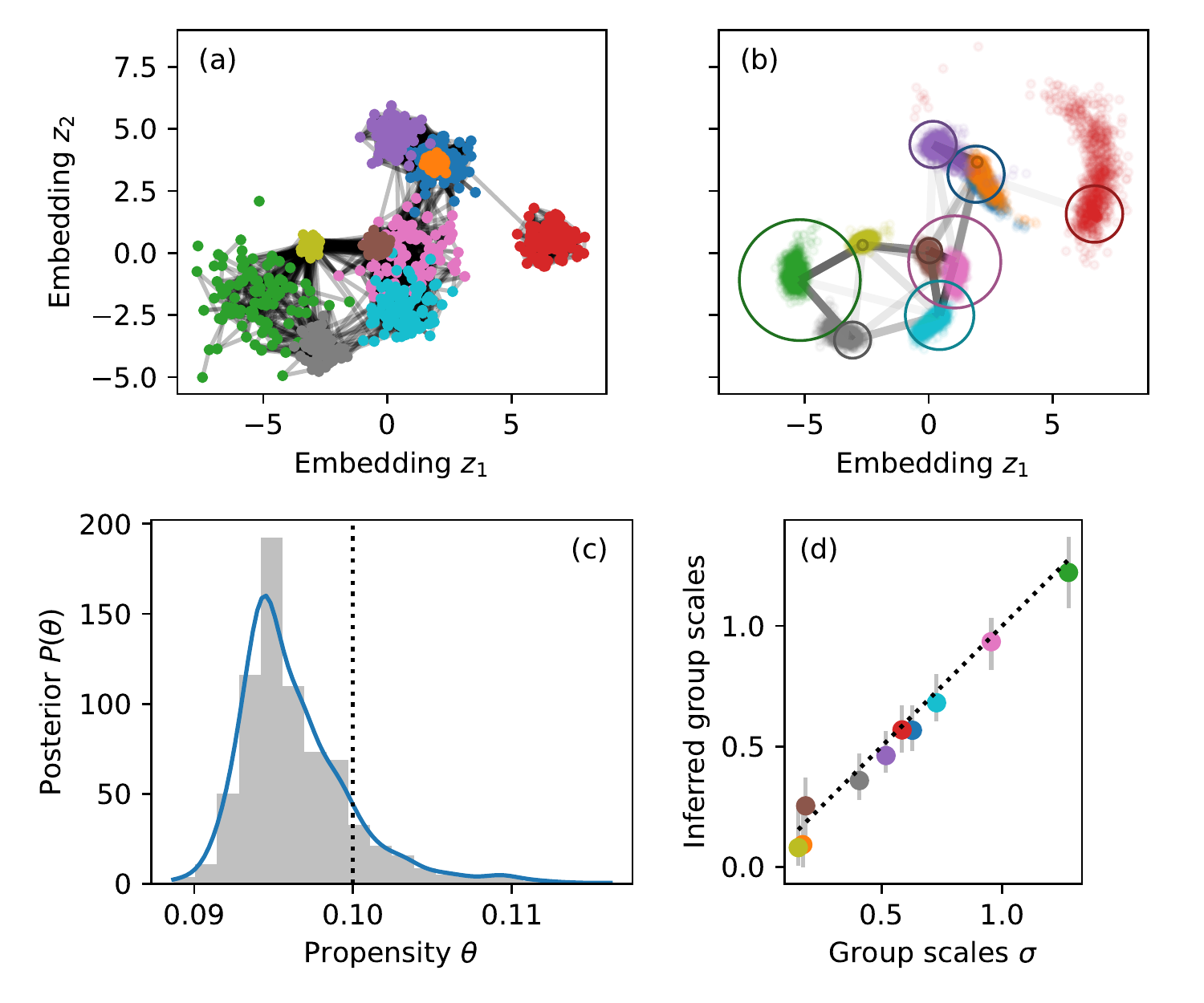}
    \caption{\emph{Aggregate network data are sufficient to infer the location and scale of clusters in a latent space model.} Panel~(a) shows synthetic data generated by the model, and each node is coloured by the cluster it belongs to. Posterior samples of the cluster centres $\mu$ inferred from only aggregate network data are shown in panel~(b). Circles centred on the maximum a posteriori estimate of each cluster location represent the posterior mean of two standard deviations $\sigma$ of the latent Gaussian clusters. Panel~(c) shows the posterior distribution of the propensity parameter $\theta$, and the vertical line represents the parameter value used to generate the data. Inferred group scales are plotted against the true group scales $\sigma$ in panel~(d). Error bars correspond to the 95\% posterior interval.}
    \label{fig:simulation}
\end{figure}

To complete the model, we impose a prior on the centres of groups
\[
    \mu \dist \Normal\left(0,\tau^2\right),
\]
where $\tau^2>0$ is the population variance. The population prior ensures that disconnected groups retain finite coordinates when the model is fit to data. We use a half-Cauchy prior for the population and group scales $\tau$ and $\sigma$, respectively~\cite{Polson2012}. The prior for kernel parameters $\theta$ will be discussed in \cref{sec:inference}. An example of a synthetic network generated by the model is shown in panel~(a) of \cref{fig:simulation}.

\subsection{Properties of aggregate network data\label{sec:properties}}

Because the individual connections $y$ and latent coordinates $z$ are not observable, we are interested in the marginal likelihood
\[
\prob{Y\mid \mu,\sigma,g,\theta}=\int dz\, \prob{Y\mid z,\theta} \prob{z\mid\mu,\sigma,g}.
\]
The distribution of $Y_{ab}$ given kernel parameters $\theta$ and node coordinates $z$ is a Poisson-binomial distribution, a generalisation of the binomial distribution allowing different probabilities for each trial. Its likelihood is difficult to evaluate, and the integral is intractable. However, we can evaluate the marginal moments of the aggregate connection volumes $Y$ which we will use to approximate the marginal likelihood. 

Taking the expectation of \cref{eq:aggregation-definition} with respect to node coordinates $z$ and connections $y$ (denoted by $\E{\cdot}_{y,z}$), we find
\begin{equation}
    \E{Y_{ab}}_{y,z}=\begin{cases}
    n_a\left(n_a-1\right)\E{\lambda_{ij}}_z&\text{if }a=b\\
    n_a n_b \E{\lambda_{ij}}_z&\text{otherwise},
    \end{cases}
    \label{eq:aggregate-mean}
\end{equation}
where $n_g$ is the number of nodes in group $g$, we have used $\lambda_{ij}=\lambda\left(z_i,z_j,\theta\right)$ for notational convenience, $z_i\dist\Normal\left(\mu_a,\sigma_a^2\right)$, and $z_j\dist\Normal\left(\mu_b,\sigma_b^2\right)$.

The second moment of within-group connections is 
\begin{equation}
    \E{Y_{aa}^2}_{y,z}=\sum_{i\neq j,k\neq l} \E{y_{ij}y_{kl}}_{y,z},\label{eq:intra-square-definition}
\end{equation}
where all nodes $i$, $j$, $k$, and $l$ belong to the same group $a$. Since the adjacency matrix elements $y_{ij}$ are independent Bernoulli random variables,
\begin{equation}
    \E{y_{ij}y_{kl}}_{y,z}=\begin{cases}
        \E{\lambda_{ij}}_z&\text{if }i=k\text{ and }j=l\\
        \E{\lambda_{ij}\lambda_{kl}}_z&\text{otherwise}.
    \end{cases}
    \label{eq:kernel-moments}
\end{equation}
Substituting into \cref{eq:intra-square-definition} yields
\begin{equation}\begin{aligned}
    \E{Y_{aa}^2}_{y,z} = n_a \left(n_a - 1\right)&\left[\E{\lambda_{ij}}_z + \E{\lambda_{ij}^2}_z + 4 (n_a - 2)\E{\lambda_{ij}\lambda_{il}}_z\right.\\
    &\quad\left.+(n_a - 2)(n_a - 3)\E{\lambda_{ij}}_z^2\right],\label{eq:intra-second-moment}
\end{aligned}\end{equation}
where repeated indices refer to the same node, distinct indices refer to different nodes, and we have used the symmetry of the kernel to simplify $\lambda_{ij}\lambda_{ji}=\lambda_{ij}^2$. The prefactors account for the number of times different terms occur (see \cref{app:statistics} for details). The variance of within-group connections is thus
\begin{align}
    \var_{y,z} Y_{aa}&=\E{Y_{aa}}_{y,z} - \E{Y_{aa}}_{y,z}^2\nonumber\\
    &=n_a \left(n_a - 1\right)\left[\E{\lambda_{ij}}_z \left(1 - \E{\lambda_{ij}}_z\right) + \var_z\lambda_{ij} + 4 (n_a - 2)\cov_z\left(\lambda_{ij},\lambda_{il}\right)\right]
    \label{eq:intra-variance}.
\end{align}
The first term accounts for variance due to individual connections, the second term captures reciprocity, and the last term encodes variance due to two edges sharing a common node. Similarly, the variance of between-group connections is
\begin{equation}\begin{aligned}
    \var_{y,z} Y_{ab} = n_a n_b &\left[\E{\lambda_{ij}}_z \left(1 - \E{\lambda_{ij}}_z\right) + (n_b - 1) \cov_z\left(\lambda_{ij},\lambda_{il}\right) \right.\\
    &\quad\left. + (n_a - 1) \cov_z\left(\lambda_{ij},\lambda_{kj}\right)\right],
    \label{eq:inter-variance}
\end{aligned}\end{equation}
where nodes $i$ and $k$ belong to group $a$ and nodes $j$ and $l$ belong to group $b$. Similar to \cref{eq:intra-variance}, the first term captures individual connection variance. The second and third terms account for variance due to two edges sharing a common node in group $a$ and $b$, respectively. In contrast to \cref{eq:intra-variance}, there is no reciprocity term because we consider directed connections from one group to another.

Having derived the marginal mean and variance for within- and between-group connections, we approximate the likelihood by independent beta-binomial distributions. Like Poisson-binomial distributions, beta-binomial distributions generalise the binomial distribution and can model overdispersion, but their likelihood is tractable. In particular,
\begin{align}
    \prob{Y_{ab}\mid\mu,\sigma,\theta}&\approx\BetaBinomial\left(Y_{ab}\mid t_{ab},\alpha_{ab},\beta_{ab}\right),\label{eq:beta-binomial-approximation}\\
    \text{where }t_{ab}&=\begin{cases}n_a(n_a-1)&\text{if }a=b\\n_a n_b&\text{otherwise}\end{cases}\nonumber
\end{align}
is the number of trials and ensures that the beta-binomial distributions have the same domain as $Y$. For notational convenience, we have omitted the explicit dependence of the shape parameters $\alpha$ and $\beta$ on $\mu$, $\sigma$, and $\theta$. We choose the parameters by matching moments with the mean in \cref{eq:aggregate-mean} and variances in \cref{eq:intra-variance,eq:inter-variance} (see \cref{app:moment-matching} for details). 

Finally, we arrive at the approximate marginal likelihood
\begin{equation}
    \prob{Y\mid \mu,\sigma,\theta}\approx\prod_{a,b=1}^r \BetaBinomial\left(Y_{ab}\mid t_{ab},\alpha_{ab},\beta_{ab}\right)
    \label{eq:approximate-likelihood}.
\end{equation}
The approximation neglects correlations between different connection volumes, but \cref{eq:approximate-likelihood} yields a sufficient approximation of the likelihood for inference in practice. These correlations are not unique to fully aggregated network data; they are also present in the partially aggregated data considered by \citet{McCormick2015}, who have similarly not modelled them.

\subsection{Gaussian connectivity kernels \label{sec:gaussian-connectivity}}

\begin{figure}
    \centering
    \includegraphics{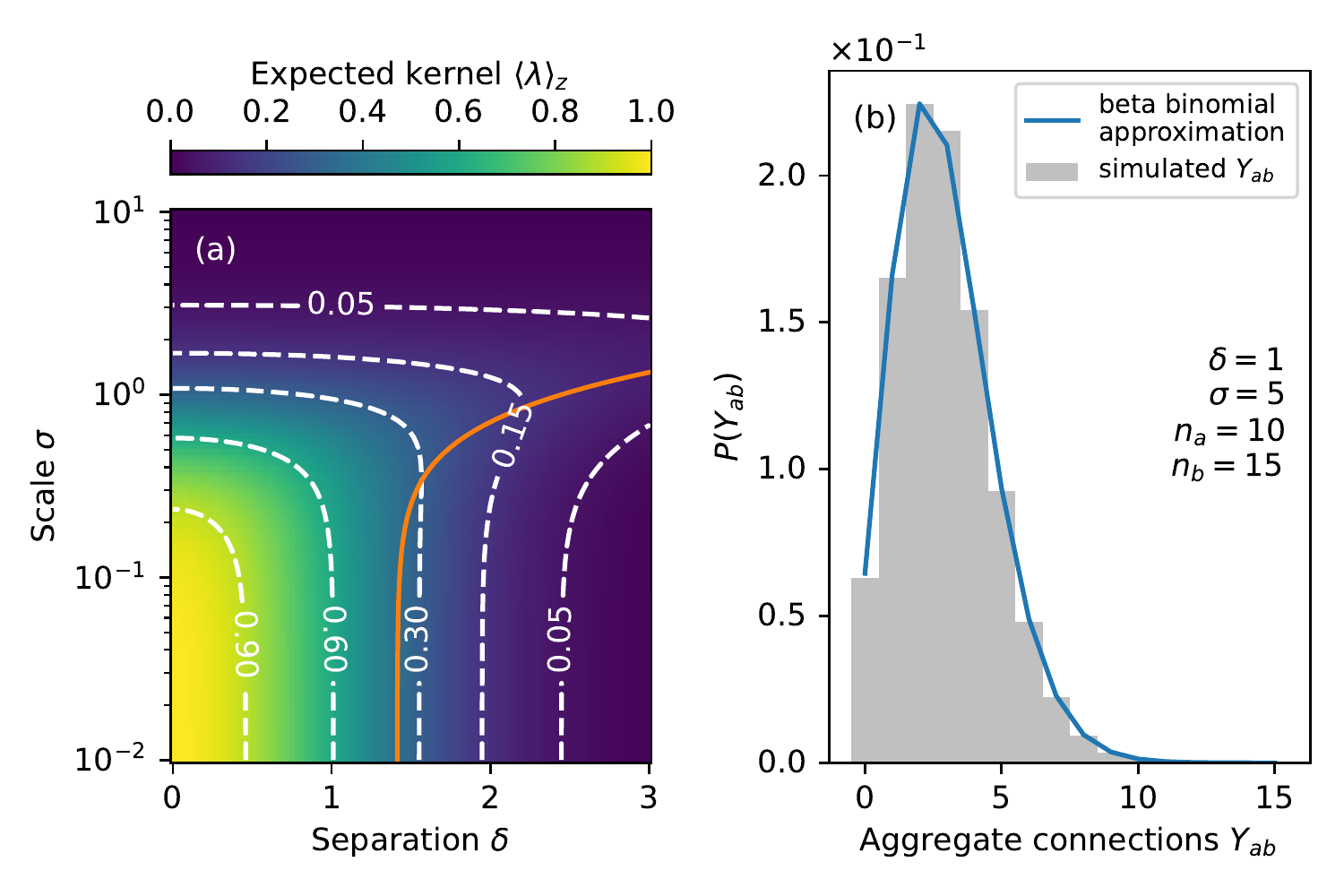}
    \caption{\emph{The expected connectivity kernel can be evaluated analytically for Gaussian kernels, and the moment-matched beta-binomial distribution approximates simulated data well.} Panel~(a) shows the expected connectivity kernel $\E{\lambda}_z$ between two groups $a$ and $b$ with the same scale $\sigma_a=\sigma_b=\sigma$ and separation $\delta=\abs{\mu_a-\mu_b}$ for a two-dimensional latent space. The solid orange curve represents the optimal scale $\sigma_{\max}$, maximising the kernel for a given separation $\delta$. Panel~(b) shows the empirical distribution (based on $10^5$ simulations) of the aggregate connection volume $Y_{ab}$ for two groups as a histogram. The two groups are separated by $\delta=1$, have shared scale $\sigma=5$, and comprise 10 and 15 nodes, respectively. The probability mass function of a moment-matched beta-binomial distribution approximates the empirical distribution well and is shown as a solid blue line.}
    \label{fig:kernel}
\end{figure}

We need to obtain expressions for the moments of the connectivity kernel on the right-hand side of \cref{eq:kernel-moments} to evaluate the approximate likelihood. Latent space models commonly use logistic kernels~\cite{Hoff2002,Handcock2007}, but their moments are intractable for the Gaussian coordinate distributions of the latent space cluster model. Following \citet{Rastelli2016}, we use a Gaussian kernel of the form
\[
    \lambda\left(x,y,\theta\right)=\theta \exp\left(-\frac{\abs{x-y}^2}{2}\right),
\]
where the propensity parameter $0\leq \theta\leq 1$ controls the overall edge density. We do not include a spatial scale parameter in the kernel because the likelihood is invariant under a global rescaling. For two nodes $i$ and $j$ respectively belonging to groups $a$ and $b$, the expected probability to connect is
\[
    \E{\lambda_{ij}}_z=\frac{\theta}{\left(1+\sigma_a^2 + \sigma_b^2\right)^{q/2}}\exp\left(-\frac{\abs{\mu_a-\mu_b}^2}{2\left(1+\sigma_a^2 + \sigma_b^2\right)}\right).
\]
A detailed derivation of the results presented in this section can be found in \cref{app:kernels}.

The expected kernel decreases exponentially with increasing separation $\delta=\abs{\mu_a-\mu_b}$ between the two clusters, as shown in \cref{fig:kernel}~(a). However, for a given separation, there is an optimal scale that maximises the expected kernel. Assuming the two clusters have a common scale $\sigma_a=\sigma_b=\sigma$, the kernel is maximised by
\[
    \sigma^2_{\max}=\frac{\max\left(\delta^2-q,0\right)}{2q}.
\]
On the one hand, when $\sigma<\sigma_{\max}$, the two clusters are isolated islands in the latent space, and the expected connection probability is small. On the other hand, when $\sigma>\sigma_{\max}$, the clusters overlap but are diffuse. The population density in the latent space is thus low, diminishing the likelihood of connections. At the optimal scale, the the two clusters overlap and the density remains sufficiently high for connections to form. When the squared distance is smaller than the dimensionality of the latent space, the connectivity is maximised by two point-mass distributions: for nonzero scales, any additional connectivity due to overlap between the distributions is thwarted by the increased separation between the peripheries.

The additional statistics required to evaluate the moments of aggregate connection volumes in \cref{eq:intra-variance,eq:inter-variance} are
\begin{align}
    \E{\lambda_{ij}^2}_z&=\frac{\theta^2}{\left(1+2\sigma_a^2 + 2\sigma_b^2\right)^{q/2}}\exp\left(-\frac{\abs{\mu_a-\mu_b}^2}{1+2\sigma_a^2 + 2\sigma_b^2}\right)\nonumber\\
    \E{\lambda_{ij}\lambda_{il}}_z&=\frac{\theta^2}{\left[(1+2\sigma_a^2+\sigma_b^2) (1+\sigma_b^2)\right]^{q/2}}\exp\left(-\frac{\abs{\mu_a-\mu_b}^2}{1+2\sigma_a^2+\sigma_b^2}\right),\label{eq:cross}
\end{align}
where node $i$ belongs to group $a$ and nodes $j\neq l$ belong to group $b$ (see \cref{app:kernels} for detailed derivations).

Having obtained expressions for the kernel moments in \cref{eq:kernel-moments}, we can evaluate the aggregate mean in \cref{eq:aggregate-mean} and variances in \cref{eq:inter-variance,eq:intra-variance}. To evaluate the quality of the moment-matched beta-binomial approximation in \cref{eq:beta-binomial-approximation}, we compared the approximation with the empirical distribution of $10^5$ independent realisations of $Y_{ab}$ obtained by simulating the data generation process at the level of individual nodes and connections. We considered two clusters with 10 and 15 nodes, respectively, shared scale $\sigma=5$, and separation $\delta=1$. The beta-binomial likelihood approximated the empirical distribution well, as shown in panel~(b) of \cref{fig:kernel}.

\subsection{Inference \label{sec:inference}}

We sought to infer the parameters (comprising cluster coordinates $\mu$, scales $\sigma$, population scale $\tau$, and propensity $\theta$) from data given only aggregate network data $Y$. To this end, we used the probabilistic programming framework Stan~\cite{Carpenter2017} to draw samples from the posterior distribution
\[
    \prob{\mu,\tau,\sigma,\theta\mid Y}\propto \prob{Y\mid\mu,\sigma,\theta}\prob{\mu\mid\tau}\prob{\tau}\prob{\sigma}\prob{\theta}.
\]
We used \cref{eq:approximate-likelihood} to approximate the first term and assumed a uniform prior on the unit interval for the propensity $\theta$. The priors for $\mu$, $\tau$, and $\sigma$ are specified in \cref{sec:model}.

The model is only identifiable up to rotations and translations in the latent space because the likelihood solely depends on distances in the space rather than absolute coordinates. Because Stan attempts to explore the entire posterior distribution, the sampler spent considerable resources exploring all possible rotations of the latent space. To reduce rotational summary, we re-parameterised the cluster centres $\mu$ such that, without loss of generality, the first $q$ centres were axis-aligned. Translational symmetry was attenuated by the population-level prior $\prob{\mu\mid\tau}$ centred at the origin. Furthermore, the propensity parameter $\theta$ was heavily correlated with the group scales $\sigma$ and population scale $\tau$ in the posterior because small propensities can be compensated for by smaller scales, increasing the density of nodes in the latent space. The presence of groups with different scales attenuated the degeneracy, but the parameters remained weakly identified. See \cref{app:parameterisation} for a more detailed discussion of the parameterisation.

Having drawn samples from the posterior, we aligned them using a rigid Procrustes transformation~\cite{Hoff2002}, i.e. samples were translated and rotated to minimise the mean squared error between samples of group locations $\mu$. Aligning the samples was necessary for common posterior summary statistics (such as posterior means or medians) to be applicable.

\section{Results\label{sec:results}}

We first considered the synthetic dataset shown in \cref{fig:simulation}~(a) that was generated by the model to validate the inference methodology. We repeated the inference ten times with different random number generator seeds and chose the fit that had the highest median posterior density. This minimised the risk of obtaining a sub-optimal fit due to local maxima of the posterior. The second-highest mode discovered by the sampler had median posterior density a factor $10^{20}$ smaller than the highest mode. Samples of the cluster centres $\mu$ reproduced the true centres well, as shown in panel~(b). The centres of clusters that were well-connected were tightly constrained in the posterior. But weakly-connected clusters, such as the red cluster on the right-hand side of panel~(b), can have a wide range of locations without affecting the likelihood significantly, giving rise to arcs of posterior samples in the latent space. Panels~(c) and~(d) show the posterior distributions of the propensity $\theta$ and group scales $\sigma$, respectively. Although their posterior distributions were consistent with the parameters that generated the data, the inferred parameters were slightly biased towards smaller values. This was a consequence of the propensity-scale degeneracy discussed in \cref{sec:inference}: the posterior density was marginally larger for samples that compress the latent space due to the half-Cauchy priors on group scales $\sigma$ and the population scale $\tau$. While it may be difficult to infer the parameters exactly, the absolute scale of the latent space is often of less interest than the relative locations of the nodes or clusters.

\begin{figure}
    \centering
    \includegraphics{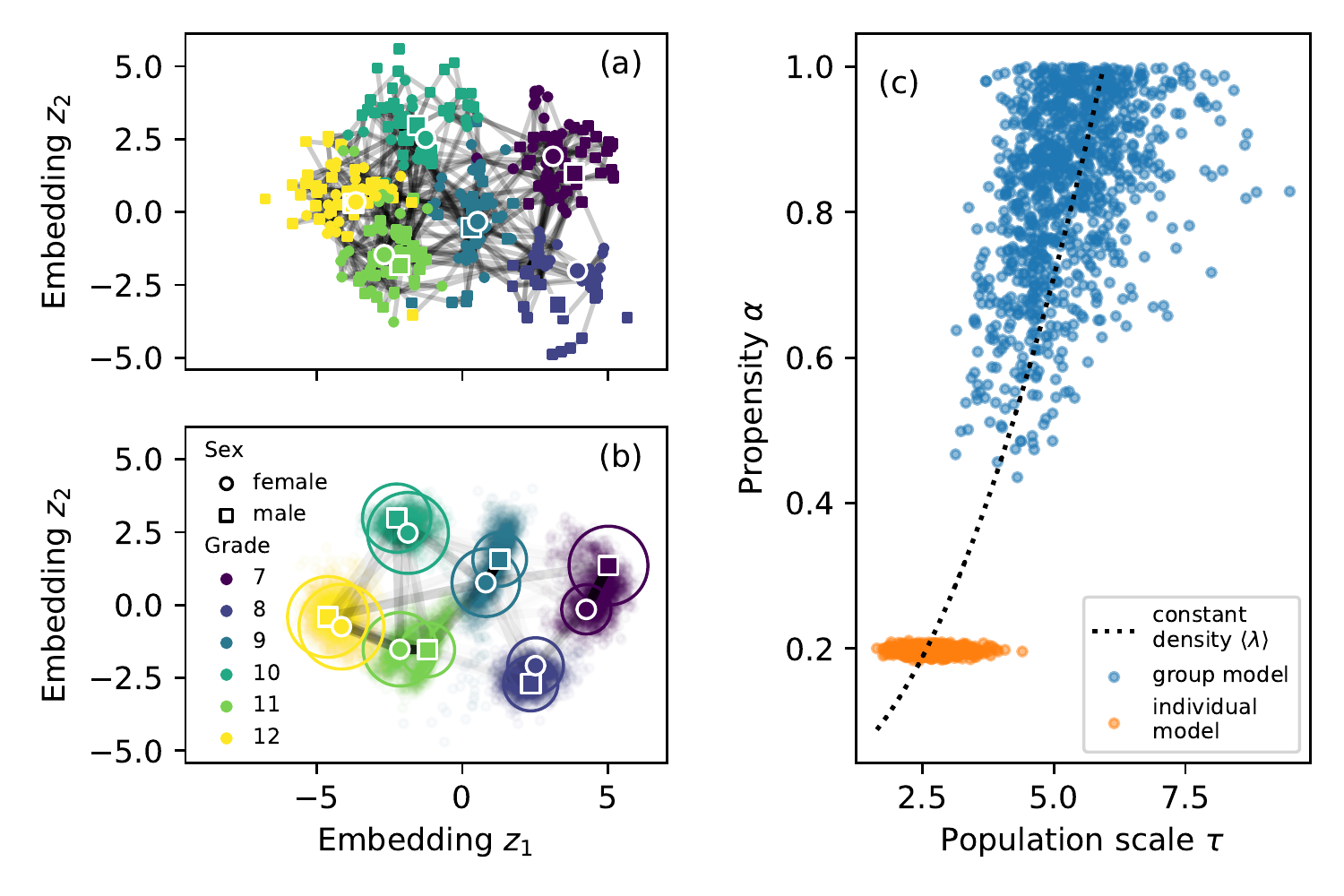}
    \caption{\emph{The latent positions of groups can be learned from both individual-level and aggregate data, although the global scale is only weakly constrained.} Panel~(a) shows the latent positions of individual students inferred using ADVI, coloured by their school grade. Cluster centres are shown with white boundaries. Posterior samples of the cluster centres $\mu$ inferred from only aggregate network data are shown in panel~(b). Circles represent the posterior mean of two standard deviations $\sigma$ of the latent Gaussian clusters. Panel~(c) shows posterior samples of the population scale $\tau$ and propensity $\theta$ for individual- and group-level fits. Both are consistent with the constant edge density contour which arises due to a degeneracy between the scale of the latent space and the propensity to form connections.}
    \label{fig:addhealth}
\end{figure}

To test the model on real-world data, we considered one of the social networks collected as part of the Adolescent to Adult Health (Add Health) study~\cite{Moody1999}. The network comprises connections between students in a community in the United States. A directed edge from $j$ to $i$ indicates that $j$ considers $i$ a friend, and the weight of the edge represents the degree of interaction, such as discussing problems or spending time together outside of school. We retained edges irrespective of their weight to obtain an unweighted network. For aggregation, groups were defined by pupils that share the same sex and school grade. We first fit a latent-space model to individual-level data to obtain a reference embedding and subsequently compared it with the embedding inferred solely based on aggregate data.

For inference given individual-level data, we used mean-field automatic differentiation variational inference (ADVI)~\cite{Kucukelbir2017} to fit the data because drawing posterior samples was prohibitively expensive. This approach approximates the posterior by a product of independent normal distributions in an unconstrained space, one for each parameter. For example, positive parameters are obtained by an exponential transform from the unconstrained space. The posterior approximation is fit by minimising the Kullback-Leibler divergence between the posterior and the approximation. We repeated the inference ten times and chose the fit with the highest evidence lower bound. Variational approximations do not suffer from non-identifiability problems in latent space models because they focus on one mode of the posterior~\cite{Bishop2006}, but mean-field approximations cannot capture correlations in the posterior. The inferred node locations are shown in \cref{fig:addhealth}~(a) together with cluster centres. As one might expect, pupils in the same grade clustered together irrespective of sex, and the positions in the latent space reflected the grade order. For the inference given only aggregate connection volumes, we drew samples from the approximate posterior in the same fashion as for the synthetic example.

The overall scale of the embeddings differed between the individual- and group-level fits, as illustrated in panel~(c). But both scales were consistent with the aggregate data because of the degeneracy between propensity and scale. Under the simplifying assumption that all node locations are drawn from a single Gaussian with scale $\tau$, the equation
\[
    \E{\lambda}=\frac{\theta}{\left(1+2\tau^2\right)^{q/2}}
\]
defines contours of equal edge density $\E{\lambda}$. Indeed, posterior samples obtained from both fits were consistent with the contour obtained by setting $\E{\lambda}$ equal to the empirical edge density of the network. Posterior samples given individual-level data did not reveal the degeneracy because mean-field variational approximations cannot capture dependencies in the posterior. To make the cluster centres $\mu$ inferred from individual connections and aggregate data comparable, we rescaled the latter by the factor $\E{\tau}_\text{individual} / \E{\tau}_\text{aggregate}$, where the expectation was taken with respect to the posteriors. The two inferences revealed similar latent structures, suggesting that aggregated connection volumes offer rich yet privacy-preserving data for studying networks.

\section{Discussion \label{sec:discussion}}

We have shown that aggregate network data are sufficient to constrain generative latent space cluster models and demonstrated the approach using both synthetic and empirical data. The method not only protects individuals' privacy (because aggregate network data are naturally privacy-preserving), but it also offers significant computational benefits (because the computational cost depends on the number of groups rather than the number of nodes). This offers the opportunity to apply generative network models to large, real-world datasets.

Students in the Add Health dataset preferentially connected with peers who are older than themselves, inducing a status order~\cite{Ball2013}. While the model can represent directed graphs, it is inherently symmetric because the expected connection volumes from group $a$ to $b$ is the same as from $b$ to $a$. Extending the model by sociability and popularity parameters could more faithfully represent such asymmetric data~\cite{Hoff2008,Hoff2021} and heavy-tailed degree distributions~\cite{Rastelli2016}. However, adding further parameters is likely to exacerbate the identifiability challenges discussed in \cref{sec:inference,sec:results}. Including higher order summary statistics, such as the number of triangles within each group or across groups, could break the degeneracy and yield improved estimates of the latent structure. Incorporating covariates, such as demographics or physical location, can help explain the observed connection patterns~\cite{Hoff2002}, and extending these ideas to aggregate network data could shed light on how people connect with one another based only on privacy-preserving data~\cite{Hoffmann2017}.

The results for the statistics of aggregate relational data in \cref{sec:properties} hold in general, but we only considered Gaussian connectivity kernels in \cref{sec:gaussian-connectivity}. This class of kernels generates strongly localised graphs because of its light tails, making the model sensitive to spurious long-range connections. Future work should explore kernels with heavier tails, such as Cauchy kernels, for more robust inference.

Of course, detailed information about individuals makes the study of social networks easier. However, from a scientific perspective, we are rarely concerned with the idiosyncratic behaviour of any \emph{one} individual but rather with the collective behaviour of members of a group. Careful modelling of aggregate network data can yield comparable insights with reduced privacy risks and computational costs, facilitating open and reproducible network analysis at population scale.

\printbibliography

\appendix

\section{Statistics of aggregate relational data\label{app:statistics}}

\begin{table}
    \centering
    \begin{tabular}{llll}
        \toprule
         & \multicolumn{2}{c}{Within $Y_{aa}^2$} & Between $Y_{ab}^2$ \\
         \cmidrule{2-3}
         Term & Directed & Undirected &  \\
         \midrule
         Prefactor      & $n_a(n_a-1)\times\ldots$ & $n_a(n_a-1)/2\times\ldots$ & $n_a n_b\times\ldots$ \\
         $y_{ij}y_{ij}$ & $1$ & $1$ & $1$\\
         $y_{ij}y_{ji}$ & $1$ & $0$ & $0$ \\
         $y_{ij}y_{il}$ & $n_a-2$ & $2(n_a-2) / 3$ & $n_b - 1$\\
         $y_{ij}y_{kj}$ & $n_a-2$ & $2(n_a-2) / 3$ & $n_a - 1$\\
         $y_{ij}y_{ki}$ & $n_a-2$ & $(n_a-2) / 3$& $0$ \\
         $y_{ij}y_{jl}$ & $n_a-2$ & $(n_a-2) / 3$& $0$ \\
         $y_{ij}y_{kl}$ & $(n_a-2)(n_a-3)$ & $(n_a-2)(n_a-3) / 2$ & $(n_a-1)(n_b-1)$ \\
         \midrule
         Total & $n_a^2(n_a-1)^2$ & $n_a^2(n_a-1)^2/4$ & $n_a^2n_b^2$\\
         \bottomrule
    \end{tabular}
    \caption{Exhaustive table of the number of occurrences of terms that contribute to the second moment of within- and between-group connections. The number of times a term occurs is equal to the column-specific prefactor multiplied by the entry in the corresponding cell.}
    \label{tbl:coefficients}
\end{table}

We provide here a more detailed derivation of the results in \cref{eq:intra-variance,eq:inter-variance}, and we extend the results to undirected and weighted networks. Recall that the second moment of the within-group connection volume in \cref{eq:intra-square-definition} is
\begin{equation}
    \E{Y_{aa}^2}_{y,z}=\sum_{i\neq j,k\neq l} \delta_{g_ia} \delta_{g_ja} \delta_{g_ka} \delta_{g_la} \E{y_{ij}y_{kl}}_{y,z}. \label{eq:intra-square-definition-recall}
\end{equation}
The sum comprises four classes of terms: first, both index pairs match, contributing $n_a(n_a-1)$ terms $\E{y_{ij}^2}_{y,z}$. Second, both pairs match but their order is reversed, contributing $n_a(n_a-1)$ terms $\E{y_{ij}y_{ji}}_{y,z}$. Third, one of the indices matches and the others differ, contribution $n_a(n_a-1)(n_a-2)$ terms each of $\E{y_{ij}y_{il}}_{y,z}$, $\E{y_{ij}y_{ki}}_{y,z}$, $\E{y_{ij}y_{kj}}_{y,z}$, and $\E{y_{ij}y_{jl}}_{y,z}$. These four terms have identical expectation because the kernel is symmetric with respect to exchange of indices. Finally, all indices differ, contributing $n_a(n_a-1)(n_a-2)(n_a-3)$ terms $\E{y_{ij}}_{y,z}^2$. It is straightforward to verify that there are $n_a(n_a-1)^2$ terms in total as expected given the double sum in \cref{eq:intra-square-definition-recall}. Collecting the terms and taking the expectation with respect to the adjacency matrix elements $y$ yields the expression in \cref{eq:intra-second-moment}. For an undirected network, we only consider indices $i<j$ and $k<l$ such that the reciprocity term vanishes and the total number of terms is reduced by a factor of four. 

The terms for between-group connections comprise $n_an_b$ terms with all indices shared, $n_an_b(n_b-1)$ terms with indices in group $a$ shared, $n_an_b(n_a-1)$ terms with indices in group $b$ shared, and $n_an_b(n_a-1)(n_b-1)$ terms with all indices distinct. A complete list of the number of contributing terms can be found in \cref{tbl:coefficients}.

The model can easily be extended to weighted networks by assuming individual connections are drawn from a Poisson distribution with rate $\lambda(z_i,z_j,\theta)$. The results above remain valid except the term with all indices matched changes from $\E{y_{ij}^2}_{y,z}=\E{\lambda_{ij}}_z$ (for Bernoulli-distributed $y$) to $\E{y_{ij}^2}_{y,z}=\E{\lambda_{ij}\left(1+\lambda_{ij}\right)}_z$ (for Poisson-distributed $y$). For weighted networks, we approximate the marginal distribution of aggregate connection volumes by a negative-binomial distribution rather than a beta-binomial distribution because its domain covers all non-negative integers.

\section{Moment matching for beta-binomial and negative-binomial distributions\label{app:moment-matching}}

Because we approximate the marginal likelihood of aggregate connection volumes by beta-binomial distributions, we need to evaluate their parameters given their moments. The mean and variance of a random variable $x$ that follows a beta-binomial distribution with $t$ trials and shape parameters $\alpha$ and $\beta$ are
\begin{align*}
    \E{x}&=\frac{t\alpha}{\alpha + \beta}\\
    \var x&=\E{x} \frac{\beta(\alpha+\beta+t)}{(\alpha+\beta)(\alpha+\beta+1)}.
\end{align*}
Assuming that the number of trials is known, we define the overdispersion factor
\[
    f = \frac{\var x}{t \rho (1-\rho)}
\]
which quantifies the dispersion of the distribution relative to a binomial distribution with constant success probability $\rho=\E{x}/t$. Then the concentration $\phi=\alpha+\beta$ of the beta-binomial distribution is
\[
\phi=\frac{t-f}{\max(f - 1, \epsilon)},
\]
where $\epsilon$ is a small constant to ensure numerical stability. In the limit $f\rightarrow 1$, we recover a binomial distribution because the concentration parameter $\phi\rightarrow\infty$. Finally, the shape parameters are $\alpha=\rho\phi$ and $\beta=(1-\rho)\phi$.

For weighted networks, we instead approximate the marginal likelihood by a negative-binomial distribution. The mean and variance of a random variable $x$ that follows a negative-binomial distribution with $t$ trials and success probability $\rho$ are
\begin{align*}
    \E{x}&=\frac{t\rho}{1 - \rho}\\
    \var x&=\frac{\E{x}}{1-\rho}.
\end{align*}
Matching moments, we find
\begin{align*}
    t&=\frac{\E{x}^2}{\max(\var x-\E{x},\epsilon)}\\
    p&= \frac{\E{x}}{\var x}.
\end{align*}
Again $\epsilon$ is a small constant to ensure numerical stability when the negative-binomial distribution closely approximates a Poisson distribution.

\section{Gaussian connectivity kernel statistics\label{app:kernels}}

For a random variable $x\dist\Normal(\mu,\sigma^2)$, 
\begin{align}
    \E{\exp\left(-\frac{x^2}{2}\right)}_x&=\frac{1}{\sqrt{2\pi\sigma^2}}\int dx\,\exp\left(-\frac{1}{2}\left[x^2+\frac{(x-\mu)^2}{\sigma^2}\right]\right)\nonumber\\
    &=\frac{\exp\left( - \frac{\mu^2}{2(1+\sigma^2)}\right)}{\sqrt{2\pi\sigma^2}} \int dx\, \exp \left(-\frac{1+\sigma^2}{2\sigma^2}\left[x - \frac{\mu}{(1+\sigma^2)}\right]^2\right) \nonumber\\
    &=\frac{1}{\sqrt{1+\sigma^2}}\exp\left(-\frac{\mu^2}{2(1+\sigma^2)}\right).\label{eq:squared-exponential-integral}
\end{align}
The second equality follows by completing the square in the exponent and the third equality by noting that the integrand is an unnormalised Gaussian density.

Assuming two nodes $i$ and $j$ belong to groups $a$ and $b$, respectively, their probability to connect is
\[
    \E{\lambda_{ij}}_z=\theta\E{\exp\left(-\frac{\abs{z_i-z_j}^2}{2}\right)}_z.
\]
Because the elements of $z_i$ and $z_j$ are normally distributed, their difference $\delta_{ij}=z_i-z_j$ is a random vector distributed as 
\[
    \delta_{ij}\dist\Normal\left(\mu_a-\mu_b,\sigma_a^2+\sigma_b^2\right).
\]
Using \cref{eq:squared-exponential-integral}, we find
\[
    \E{\lambda_{ij}}_z=\frac{\theta}{\left(1+\sigma_a^2+\sigma_b^2\right)^{q/2}}\exp\left(-\frac{\abs{\mu_a-\mu_b}^2}{2(1+\sigma_a^2+\sigma_b^2)}\right).
\]

For the second moment of the connectivity kernel, we have
\begin{align*}
    \E{\lambda_{ij}^2}_z&=\theta^2\E{\exp\left(-\abs{z_i-z_j}^2\right)}_z\\
    &=\frac{\theta^2}{\left(1+2\sigma_a^2+2\sigma_b^2\right)^{q/2}}\exp\left(-\frac{\abs{\mu_a-\mu_b}^2}{1+2\sigma_a^2+2\sigma_b^2}\right),
\end{align*}
where we have again employed \cref{eq:squared-exponential-integral}.

We finally consider the cross term
\[
    \E{\lambda_{ij}\lambda_{il}}_z=\theta^2\E{\exp\left(-\frac{\abs{z_i-z_j}^2+\abs{z_i-z_l}^2}{2}\right)}_z,
\]
where $i$ belongs to group $a$ and $j$ and $l$ are members of $b$. \Cref{eq:squared-exponential-integral} cannot be applied directly because the squared terms share the common variable $z_i$. We thus define the auxiliary variables
\begin{align*}
    \xi&=\frac{2z_i - z_j - z_l}{\sqrt 2}\sim\Normal\left(\sqrt 2\left(\mu_a-\mu_b\right),2\sigma_a^2+\sigma_b^2\right)\\
    \chi&=\frac{z_j-z_l}{\sqrt 2}\sim\Normal\left(0,\sigma_b^2\right)
\end{align*}
and note that $\abs{z_i-z_j}^2+\abs{z_i-z_l}^2=\abs{\xi}^2+\abs{\chi}^2$. The auxiliary variables are uncorrelated, and applying \cref{eq:squared-exponential-integral} yields \cref{eq:cross}:
\[
    \E{\lambda_{ij}\lambda_{il}}_z=\frac{\theta^2}{\left[\left(1+2\sigma_a^2+2\sigma_b^2\right)\left(1+\sigma_b\right)\right]^{q/2}} \exp\left(-\frac{\abs{\mu_a-\mu_b}^2}{1+2\sigma_a^2+\sigma_b^2}\right).
\]

\section{Model parameterisation for efficient sampling\label{app:parameterisation}}

We consider two changes to the parameterisation of the model to aid the sampling process. First, the likelihood is invariant with respect to global translations and rotations, and exploring all possible configurations of the posterior is prohibitively expensive. We thus parameterise the cluster centres as
\[
    \mu_{as}=\gamma_{as} + \nu_s,
\]
where $\nu$ is a $q$-dimensional translation vector, $s$ indexes the dimensions of the latent space, and $\gamma$ is a strictly lower diagonal $r$ by $q$ matrix, i.e. $\gamma_{as}=0$ if $a\leq s$. The $q (q + 1) / 2$ zeroes eliminate global rotational symmetry, but they also pin the first cluster at the origin which can give rise to tension with the population prior $\mu\dist\Normal(0,\tau^2)$ because the ordering is arbitrary. For example, suppose the first cluster is not connected to the rest of the network. It should thus be located in the periphery of the latent space rather than at the origin. The translation vector $\nu$ provides the flexibility to align the ``centre'' of the network with the origin. This formulation is equivalent to setting $\mu=\gamma$ and inferring the location of the population prior $\mu\dist\Normal(\nu,\tau^2)$ with an improper flat prior on $\nu$.

Second, the propensity $\theta$ and group scales $\sigma$ have a complex dependence in the posterior, as illustrated by \cref{fig:addhealth}~(c). In particular, the within-group edge density for a cluster with scale $\sigma$ is
\[
    \E{\lambda}_z = \frac{\theta}{\left(1+2\sigma^2\right)^{q/2}}.
\]
We change variables to $\eta=\left(1+2\sigma^2\right)^{-q/2}$ such that $\E{\lambda}_z=\theta\eta$; $\eta$ is the fraction of the maximal edge density a cluster can realise given the propensity $\theta$. The reparameterisation does not eliminate the degeneracy but simplifies the shape of the posterior. The scale and the associated Jacobian, which is needed to account for the change of variables, are
\begin{align*}
    \sigma(\eta) &=\sqrt{\frac{\eta^{-2/q} - 1}{2}}\\
    \abs{\frac{d\sigma}{d\eta}}&=\frac{\eta ^ {- 2 / q - 1}}{2 q \sigma}.
\end{align*}

\end{document}